\documentstyle[prb,aps,epsf]{revtex}
\begin{document}
\draft
\title{\bf Band structures 
 of  periodic carbon nanotube junctions and their symmetries analyzed by
 the effective mass approximation}
\author{Ryo Tamura and Masaru Tsukada }
\address{Department of Physics, Graduate School of Science, University
  of Tokyo, Hongo 7-3-1, Bunkyo-ku, Tokyo 113, Japan}
\maketitle
\begin{abstract}
The band structures of the periodic nanotube junctions are investigated
 by the effective mass theory and the tight binding model. 
 The periodic junctions are constructed by introducing pairs of
 a pentagonal defect and a heptagonal defect periodically in the
 carbon nanotube. 
 We treat the periodic junctions 
 whose unit cell is  composed by two kinds of 
 metallic nanotubes with almost same 
 radii, the ratio of which is between 0.7 and 1 .
 The discussed energy region is near the undoped Fermi level
 where the channel number is kept to two, so there are two bands.
 The energy bands are expressed with closed
 analytical forms by the effective
 mass theory with some assumptions, and they coincide well with 
 the numerical results
  by the tight binding model.
 Differences between the two methods are also discussed.
 Origin of correspondence between the band structures and the phason pattern
 discussed in Phys. Rev. B {\bf 53}, 2114, is clarified.
 The width of the gap and the band are in inverse
 proportion to the length of the unit cell, which is the sum
 of the lengths measured
 along the tube axis in each tube part and along 'radial' direction
 in the junction part.
 The degeneracy and repulsion between the two bands are determined
 only from symmetries.
\end{abstract}
\pacs{72.80.Rj,73.20.Dx,72.10.Fk}
\twocolumn
\narrowtext
\section{Introduction}
 Carbon nanotubes are one-dimensional  structures formed
 by rolling up the honeycomb lattice of the monolayer graphite.\cite{tube}
 Their radius and  length are of nanometer and micrometer sizes, respectively.
 One of their interesting features theoretically predicted \cite{tubetheory}
 and investigated experimentally \cite{experiment}
 is that they become metallic or 
 semiconducting according to the radius and the helicity 
 of the honeycomb lattice forming  the tubes.
 Especially metallic nanotubes are expected to be used
 as electric leads with nanometer size.
 Thus we concentrate our discussion to the metallic nanotubes
 in this paper.

 A junction connecting different nanotubes can be formed
 without dangling bonds by introducing a pair of
a pentagonal defect and a heptagonal defect.
\cite{iijimajunction,tamurajunction1,tamurajunction2,saitojunction,chico}
 Such defects are called disclinations and are necessitated to form
 various  structures composed of curved surface of graphitic layer,\cite{discl,discl2} for example, fullerenes,\cite{c60} minimal-surface structures \cite{minimal},
 torus structures\cite{ihara}, cap structures at the end of the tube \cite{cap}
 and helical nanotubes \cite{akagi,helix}.
The tight binding calculations show that 
 the electronic states near the undoped Fermi level are drastically
 changed according to whether the Kekule patterns started from
 the different disclinations  match with each other. \cite{discl,akagi}
 The boundary between the mismatched Kekule patterns is called
  'phason line' in Ref.\cite{akagi}
 and the same word is used also in this paper.
 The Kekule pattern represents the periodicity of the Bloch states
 at the  $K $ and  $K' $ corner points in the 2D Brillouin zone of the
 monolayer graphite, at which the Fermi level exists.
 It means that certain characters of the Bloch states remain,
 though the Bloch states themselves can not be the eigen states of
 the systems. In fact the electronic states of the junctions
 near the Fermi level can be described fairly well by the Bloch states
 multiplied by envelop functions as shown in this paper.

The way of tiling the honeycomb pattern on the nanotubes determines
 whether they are metallic or semiconducting.
  Adding only one atomic raw to a nanotube can change 
 a metallic nanotube into a semiconducting one or vice versa.
 But once one knows whether the nanotube is semiconducting or metallic,
necessary information to determine the electronic structures near 
 the Fermi level is  only about the size and the shape, i.e.,
  one can forget detailed information about the honeycomb lattice such as
  the direction of the honeycomb row with respect to the tube axis.
For example, the gap of the semiconducting tube is in inverse proportion
 to its radius  because of the isotropic linear dispersion relation at the  $K $ and  $K' $ corner points.
Another example is the conductances of the junction connecting
 two metallic nanotubes.
It is  determined almost only by the ratio of the circumferences
 of the nanotubes
 and the ratio  $|E/E_c| $, where  $E_c $ is threshold energy above which
 the channel number increases.\cite{tamurajunction1,tamurajunction2}
Both examples are independent of the 
helicity of the honeycomb lattice and suggest
 that some continuum theory ignoring the atomic details but
 including only the size and shape of the systems
is effective.
The continuum theory describing the envelop factors of the wave functions
 is known as
 the effective mass theory or the  $k \cdot p $ approximation.
Purpose of this paper is to explain the origin of 
the correspondence between the band structures and
 the phason lines of the periodic nanotube junctions 
by using this theory.\cite{akagi}\cite{helixnote}
Especially, whether the bands are degenerate or
 avoid each other is related to the symmetries as discussed in section 
 \ref{symmetry}.
 Furthermore dependence of the band structures
 on the size and the shape of the system is discussed
 in detail based on this theory.

\section{Effective mass theory and its application to the single nanotube
 junction}
 The single junction is discussed in this section, which provides a basis for
 the discussions of the periodic junctions.
 The expression is changed from that of our previous paper \cite{pre}
 and other references \cite{ajiki}\cite{matsumura} to facilitate discussions.
First of all, we explain  the  Bloch state of  a monolayer graphite
 forming the single wall nanotube
by the tight binding model
 and relate it to the effective mass theory.\cite{ajiki}
Fig. \ref{tubetenkai}  shows the development map of the nanotube.
The vector  $\vec{R} $ represents the circumference of the tube.
Two  parallel lines  perpendicular to  $\vec{R} $ and parallel
 to the tube axis are connected with each other to form the tube.
Here we use two pairs of the vectors  $\{ \vec{e}_1,\vec{e}_2 \} $
and  $\{ \vec{e}_{x}, \vec{e}_{y} \} $, where  $\vec{e}_{x}=(\vec{e}_1+\vec{e}_2)/\sqrt{3} $ and  $\vec{e}_{y}=\vec{e}_2-\vec{e}_1 $, to represent components of vectors on the development map.
 For example, the components of  $\vec{R} $  in Fig. \ref{tubetenkai} are represented as   $(R_1,R_2)=(2,5) $ and  $(R_{x},R_{y})= (7\sqrt{3}/2,1.5) $.
In this paper, we concentrate our discussion to the metallic nanotube, 
 so that only the tube of which  $R_1-R_2 $ is an integer  multiple of three is considered.
\cite{tubetheory}
The four  vectors
have the same length which is about 0.25 nm and
 denoted by  $a$ hereafter.
The amplitudes of the wave function at $\vec{q}=(q_1,q_2)$ and 
 at $\vec{q}+\vec{\tau}=(q_1+\frac{1}{3},q_2+\frac{1}{3})$
are represented by
 $\psi_A(\vec{q})$ and $\psi_B(\vec{q})$, respectively,
 with integer components $q_1$ and $q_2$.
The wave function $\psi$ can be represented 
by the Bloch state as $\psi_i(\vec{q})= e^{i(k_1 q_1+k_2 q_2)a}
\psi_i(0)\;\;\;(i=A,B)$.  
When the metallic nanotube is not doped,  i.e.  ,the $\pi$ band is half filled,
 the Fermi energy locates 
at the $K$ and $K'$ corner points in the 2D Brillouin zone:
 the corresponding wave numbers $(k_1a,k_2a)$ are $(2\pi/3,-2\pi/3)$ and
$(-2\pi/3,2\pi/3)$, respectively.
The corresponding energy  position,  i.e. , the Fermi level of the undoped
 system, is taken to be zero hereafter.
Near the undoped Fermi level, i.e., 
 when the wavenumber $\vec{k}$ is near the  corner
point K, the wavenumber $\vec{k'}$ measured from the $K$ point,
$(k'_1a,k'_2a)=(k_1a-2\pi/3,k_2a+2\pi/3)$, is small so that
the phase factors can be linearized as
 $\exp (ik_1a)=w \exp (ik'_1a) \simeq w(1+ik'_1a)$ and  $\exp (ik_2a)=w^{-1}
\exp (ik'_2a) \simeq w^{-1}(1+ik'_2a)$ , where $w \equiv \exp (i2\pi/3)$.
Then  Schr$\ddot{\rm{o}}$dinger 
 equation of a simple 
 tight binding model for the Bloch state becomes

\begin{equation}
E \psi_A(\vec{q})=\frac{\sqrt{3}}{2}\gamma a(k'_{y}+ik'_{x})
\psi_B(\vec{q})\;\;,
\label{blochA}
\end{equation}
and
\begin{equation}
E \psi_B(\vec{q})=\frac{\sqrt{3}}{2}\gamma a(k'_{y}-ik'_{x})\psi_A(\vec{q})\;\;,
\label{blochB}
\end{equation}
where $k'_{x}=(k'_1+k'_2)/\sqrt{3}$ and $k'_{y}=k'_2-k'_1$.
Here $\gamma ( \simeq -2.7$eV) is the hopping integral between the nearest neighboring sites.
In this tight binding model, only the $\pi$ orbital is considered and mixing between the $\sigma$ and  the $\pi$ orbital caused by the finite curvature is neglected.
The solution of these equations shows the linear dispersion relation,
\begin{equation} |\vec{k'}|=\pm \frac{2E}{\sqrt{3}\gamma a}\;\;.
\label{dispersion}
\end{equation}
For the one-dimensional band which intersects the $K$ point,
the periodic boundary condition around the circumference is
$R_1 k'_1+R_2 k'_2=0$.
From this condition, one can show that phase difference 
between $A$ sublattice and $B$ sublattice is represented by
\begin{equation}
\psi_B(\vec{q})/\psi_A(\vec{q})=\pm   \exp (i\eta)\;\;,
\label{abphase}
\end{equation}
where $\eta$ is the angle of $\vec{R}$ with respect to $\vec{e}_{x}$ measured  anti-clockwise as shown Fig. \ref{tubetenkai} .

In the effective mass theory, the wave function is represented  by
\begin{equation}
\psi_{i}(\vec{q})=F_i^{K}(\vec{q})w^{(q_1-q_2)}
 +F_i^{K'}(\vec{q})w^{(q_2-q_1)}\;\;\;(i=A,B).
\label{defFA}
\end{equation}
where $F_{A,B}^{K,K'}$, $w^{(q_1-q_2)}$, $w^{(q_2-q_1)}$  are the
 envelop wave functions, the Bloch state wave function at the $K$ point and
 that at the $K'$ point, respectively.
This definition of F's is different from
 that of our previous paper \cite{pre} and other references \cite{ajiki,matsumura} by certain factors.
The reason why this definition is used is that the  representation
 of the time reversal operation $I$ becomes simpler as
\begin{equation}
I(F_A^K,F_B^K,F_A^{K'},F_B^{K'})=((F_A^{K'})^*,(F_B^{K'})^*,(F_A^K)^*,(F_B^K)^*)\;\;,
\label{timerev}
\end{equation}
so that the corresponding scattering matrix becomes symmetric,
 which is important in the discussions in section \ref{symmetry}.

When the Fermi level $E_F$ is close to zero,
spatial variance
 of the envelop function is slow compared to the lattice constant, $a$,
 so that it is a good approximation to take only the first order term in
 the Taylor expansion of the envelop function.
This approximation corresponds to 
the replacement  $k'_{x} \rightarrow -i \partial_{x}$ and $k'_{y} \rightarrow -i \partial_{y}$  in eq. (\ref{blochA}) and eq. (\ref{blochB}),
  from which one obtains
\begin{equation}
(-i\partial_{y}+ \partial_{x})F^K_B= k F^K_A
\label{kpKB}
\end{equation}
\begin{equation}
(-i\partial_{y}- \partial_{x})F^K_A= k F^K_B\;\;.
\label{kpKA}
\end{equation}
Here $k \equiv |\vec{k'}|$ and the isotropic linear dispersion relation
(\ref{dispersion}) is used.
The equations of the envelop wave functions $F_{A,B}^{K'}$ for
 the $K'$ corner point can be easily obtained in a similar way as
\begin{equation}
(i\partial_{y}+ \partial_{x})F^{K'}_B= k F^{K'}_A
\label{kpK'B}
\end{equation}
\begin{equation}
(i\partial_{y}- \partial_{x})F^{K'}_A= k F^{K'}_B\;\;.
\label{kpK'A}
\end{equation}
Hereafter the envelop wave functions $F$'s 
are often simply called the wave functions.

It can be seen from eq. (\ref{abphase}) that the Bloch state wave function
 for the one dimensional band intersecting the $K$ point
is represented by the envelop wave functions $(F_A^K, F_B^K,F_A^{K'},F_B^{K'})$
 as
\begin{equation}
\psi_{K\pm}=(e^{-i\eta/2}, 
\pm e^{i\eta/2},0,0 ) e^{\pm i(\vec{k'} \cdot \vec{q})}
\;\;,
\label{propK}
\end{equation}
 where the upper sign and the lower sign
 represent the direction of the propagating waves.
 Here, positive direction is taken to be from the thicker tube
 to the thinner tube, as is shown in Fig. \ref{junctiontenkai}.
 From the propagating waves near the $K$ point,
 the other propagating waves $\psi_{K'\pm}$ 
 are obtained by the time reversal operation (\ref{timerev})
 as
\begin{equation}
\psi_{K'\pm}=I \psi_{K\mp}=(0,0,e^{i\eta/2}, 
\mp e^{-i\eta/2}) e^{\pm i(\vec{k'} \cdot \vec{q})}
\;\;.
\label{propK'}
\end{equation}
Note that the direction of the propagation is reversed by the time reversal
 operation $I$.

When the direction of $x'=x^{(j)};\; (j=5,7)$
 is taken to be parallel
to the circumference of each tube  as shown Fig. \ref{junctiontenkai},
$k'_{x'}$ is quantized as
$k'_{x'}(n)=2\pi n/R$ and $k'_{y'}$ is given by $k'_{y'}(n)=\sqrt{k^2-k_{x'}(n)^2}$.
Here $n$ is an integer representing a number of nodes around the
 circumference and $R$ is the circumference of the tube.
When $k'_{y'}(n)$ is a real number, the channel $n$ is open and
 the corresponding wave function is extended, 
 otherwise the channel is closed and the wave function shows exponential
 grow or decay.
The number of the open channel is called the channel number.
When the Fermi energy is zero,
 only the channel $n=0$ is open, and therefore the channel number is 
 kept  to two irrespective of $R$.

The electronic states at the  Fermi energy 
($E_F=0$) govern the electron transport for the undoped system,
so discussion in this paper is concentrated to  the energy region
 near zero where the channel number is kept to two.
In order to discuss the wave function in the junction
part, the polar coordinate $(r,\theta)$ is useful.
Its relation to the coordinate $(x,y)$ is the 
usual one, i.e. ,$ r=\sqrt{x^2+y^2}$, $\tan \theta =y/x$.
Fig. \ref{junctiontenkai} is the development map of the nanotube junction where
 the origin of the coordinate  is defined.\cite{saitojunction}
A heptagonal defect and a pentagonal defect are introduced
at $P_5(=Q_5)$ and $P_7(=Q_7)$, respectively.
 Thus from now on, the indices '7' and '5' are
 used to represent the thinner and the thicker tube, respectively.
The equilateral triangles '$\Delta  OP_7Q_7$'
 and '$\Delta OP_5Q_5$'
  with bases '$P_7Q_7$' and '$P_5Q_5$'
 have common apex $ O$,
 which is taken to be the origin of the coordinate
$(x,y)$ in this paper. \cite{saitojunction}
Then the wave function satisfies the wave equation $(z^2\partial_z^2+z\partial_z+\partial_{\theta}^2+z^2) F = 0$, where $z=kr$.
The solution is represented by Bessel functions $J_m$ and Neumann functions
 $N_m$  as 
\begin{equation}
F= \sum_{m=-\infty}^{\infty} e^{im\theta}(c_m J_{|m|}(z)+d_m N_{|m|}(z)) \;\;.
\label{JmNm}
\end{equation}

The boundary condition  in the junction part is 
$ \psi(r,\theta+\pi/3)= \psi(r,\theta)$.
It is represented by
\begin{eqnarray}
\psi_A(q_1,q_2)= \psi_B(q_1+q_2,-q_1-1) \nonumber \\
\psi_B(q_1,q_2)= \psi_A(q_1+q_2+1,-q_1-1) 
\label{junctbound}
\end{eqnarray}
where $\vec{q}=(q_1,q_2)$ is the position of arbitrary atoms 
 in the junction part.
Fig. \ref{rotbound} shows an example where $q_1=q_2$, i.e.,  the points 
$B(2i,-i-1)$ and $A(2i+1,-i-1)$
on line $OQ$ are transformed into  $A(i,i)$ and $B(i,i)$ on line $OP$, respectively,
under the rotation by $\pi/3$ with respect to the origin $O$.
The angle $\theta$, i.e., the direction of $OQ$ can be taken arbitrary. 
 In Fig. \ref{rotbound}, 
 it is taken to be parallel to the bond, i.e., $\theta=-\pi/3$, 
 only to simplify the presentation.

From eq. (\ref{junctbound}) and eq. (\ref{defFA}),
the boundary conditions in the junction part are represented by
\begin{equation}
F^{K'}_A(z,\theta+\pi/3)=wF^K_B(z,\theta)\;\;,
\label{bound1}
\end{equation}
\begin{equation}
F^{K}_B(z,\theta+\pi/3)=wF^{K'}_A(z,\theta)\;\;,
\label{bound2}
\end{equation}
\begin{equation}
F^{K}_A(z,\theta+\pi/3)=\frac{1}{w}F^{K'}_B(z,\theta)\;\;,
\label{bound3}
\end{equation}
\begin{equation}
F^{K'}_B(z,\theta+\pi/3)=\frac{1}{w}F^{K}_A(z,\theta) \;\;.
\label{bound4}
\end{equation}
 Similar boundary conditions are discussed by
 Matsumura and Ando.\cite{matsumura}
 But  the boundary conditions, eq.  (\ref{bound1}), eq.  (\ref{bound2}),
 eq.  (\ref{bound3})  and eq.  (\ref{bound4})
 are different from those of Matsumura and
 Ando by certain factors due to the difference of the definition of 
 $F_{A,B}^{K,K'}$.
In these equations and hereafter, $w \equiv e^{i2\pi/3}$.
From eq. (\ref{bound1}) and eq. (\ref{bound2}), terms in eq. (\ref{JmNm})
for $F^{K'}_A$ and $F^K_B$ are not zero only when $m=3 l+2 \;\;(l=$integer
).
 Because the open channel $n=0$ has no node along the circumference,
 it is better fitted to the components with smaller $|m|$ in eq. (\ref{JmNm}) 
 than to those with larger $|m|$.
So we assume that one can neglect all the terms except those with $l=0$ and $l=-1$ in eq. (\ref{JmNm}) (Assumption I). Then the wave functions can be written as
\begin{equation}
F^{K'}_A=e^{2i\theta} f_2(z)+e^{-i\theta} f_1(z),
\label{fk'a}
\end{equation}

and

\begin{equation}
F^{K}_B=e^{2i\theta }f_2(z)-e^{-i\theta} f_1(z)\;\;,
\label{fkb}
\end{equation}

where

\begin{equation}
f_m(z)=c_mJ_m(z)+d_mN_m(z) \;\;(m=1,2).
\label{fm}
\end{equation}

From eq. (\ref{kpKA}) and eq. (\ref{kpK'B}), the other two wave functions $F^{K'}_B$ and $F^K_A$ can be derived from $F^{K'}_A$ and $F^K_B$ as

\begin{equation}
F^{K'}_B=-e^{i\theta} \tilde{f}_2(z)+e^{-i2\theta} \tilde{f}_1(z) \;\;,
\label{fk'b}
\end{equation}

\begin{equation}
F^{K}_A=e^{i\theta} \tilde{f}_2(z)+e
^{-i2\theta} \tilde{f}_1(z) \;\;,
\label{fka}
\end{equation}

where

\begin{eqnarray}
\tilde{f}_1(z)=c_1J_2(z)+d_1N_2(z),\nonumber \\
\tilde{f}_2(z)=c_2J_1(z)+d_2N_1(z)\;\;,
\label{tildef}
\end{eqnarray}
by using the recursion formula of the Bessel functions and Neumann
 functions.
It is easily confirmed that eq. (\ref{fk'b}) and eq. (\ref{fka}) satisfy
 the boundary conditions eq. (\ref{bound3}) and eq. (\ref{bound4}).
 The amplitude of the open channel in the tube, which is denoted by $\alpha$,  
 is obtained from eq. (\ref{abphase}) as
\begin{equation}
\alpha_{j\pm}^K=\frac{1}{\sqrt{2R_j}}\int_{Q_j}^{P_j} dx^{(j)} 
(e^{i\frac{\eta_j}{2}}F^K_A \pm  e^{-i\frac{\eta_j}{2}}F_B^K)\;\;(j=5,7)\;\;,
\label{alpha}
\end{equation}
for the $K$ point.
The indices $+$ and $-$ mean directions in which the electronic waves
propagate.
$R_5$ and $R_7$ are the circumferences of
 the thicker tube and thinner tube. Path of integral is the straight
 line $P_jQ_j$, the angle of which is denoted by $\eta_j$  with respect to $x$ axis.
Equations for $\alpha^{K'}_{j\pm}$  are obtained from eq. (\ref{alpha}) by replacing $\pm$  and $K$ in the r. h. s with $\mp$ and $K'$, respectively.
To simplify the calculation, the integrations
in the above equations are transformed as

\begin{equation}\int_{Q_j}^{P_j} dx^{(j)} \rightarrow R_j \int_{-\frac{2}{3}\pi+\eta_j}
^ {-\frac{\pi}{3}+\eta_j}  d\theta\;\;.
\end{equation}

If variation of the wave function  along the radial directions
 is slow near $r=R_j$, this replacement can be allowed (Assumption II).
The relation between the amplitudes of the open channel in each tube,
$\vec{\alpha}_j=\; ^t\,(\alpha_{j+}^K,\alpha_{j+}^{K'},\alpha_{j-}^{K'},
\alpha_{j-}^{K} )$, and the coefficients representing
 the wave functions in the junction part,
 $\vec{c}=\; ^t\,(c_2,d_2,c_1,d_1 )$, are summarized in the followings.

\begin{equation}
\vec{\alpha}_j=\sqrt{R_j}P(\eta_j) M \Lambda(\eta_j) L(k R_j ) \vec{c}\;\;,
\label{a5mat}
\end{equation}
where $M$ is a constant matrix given by
\begin{equation}
M=
\left( \begin{array}{cccc}
-i& 0 & 0 & -\frac{\sqrt{3}}{2}\\
0 & -\frac{\sqrt{3}}{2} & -i& 0 \\
0 &  -\frac{\sqrt{3}}{2}& i & 0 \\
i & 0 & 0 &  -\frac{\sqrt{3}}{2}
\end{array} \right)\;\;.
\label{M}
\end{equation}
$\Lambda(\eta)$ is a diagonal matrix, where $\Lambda_{1,1}=\Lambda_{3,3}^*=e^{i \eta}$ and $\Lambda_{2,2}=\Lambda_{4,4}^*=e^{2i \eta}$.
$P(\eta)$ is defined by eq. (\ref{alpha}) as 
\begin{equation}
P(\eta) =
\left( \begin{array}{cccc}
e^{i\frac{\eta}{2}},& e^{-i\frac{\eta}{2}}, &0, &0\\
0, & 0, &e^{-i\frac{\eta}{2}},& -e^{i\frac{\eta}{2}} \\
 0, & 0,& e^{-i\frac{\eta}{2}},& e^{i\frac{\eta}{2}} \\
e^{i\frac{\eta}{2}},& -e^{-i\frac{\eta}{2}}, &0, &0
\end{array} \right)\;\;.
\label{P}
\end{equation}
 The matrix elements of  $L(z)$ are $L_{11}=L_{33}=J_1(z)$, $L_{12}=L_{34}=
N_1(z)$, $L_{21}=L_{43}=J_2(z)$ and $L_{22}=L_{44}=N_2(z)$. The other matrix
 elements of $L(z)$ are zero.
From eq. (\ref{a5mat}), the relation 
  between $\vec{\alpha}_7$ and $\vec{\alpha}_5$
 is represented by
  three parameters,  $\beta \equiv R_7/R_5$ (ratio between the radii)
  ,$\phi \equiv \eta_7-\eta_5$ (angle between the tube axes
 in the development map) and $z \equiv k R_5$ as
\begin{equation}
\left( \begin{array}{c}
\vec{\alpha}_{7+} \\
\vec{\alpha}_{7-}\\
\end{array} \right)
 =
\left( \begin{array}{cc}
t_{1} ,& t_{2}^* \\
t_{2} ,& t_{1}^* \\
\end{array} \right)
\left( \begin{array}{c}
\vec{\alpha}_{5+} \\
\vec{\alpha}_{5-}\\
\end{array} \right)\;\;.
\label{transfer}
\end{equation}
where
\begin{equation}
t_1 =h_+
\left( \begin{array}{cc}
\cos(\frac{3}{2}\phi),& i \sin(\frac{3}{2}\phi) \\
i\sin(\frac{3}{2}\phi),& \cos(\frac{3}{2}\phi) \\
\end{array} \right)
\label{t1}
\end{equation}

\begin{equation}
t_2 =h_-
\left( \begin{array}{cc}
-\cos(\frac{3}{2}\phi),& -i \sin(\frac{3}{2}\phi)\\
i\sin(\frac{3}{2}\phi),& \cos(\frac{3}{2}\phi) \\
\end{array} \right)\;\;.
\label{t2}
\end{equation}
The factors $h_+$ and $h_-$ are represented by
\begin{equation}
h_{\pm}= -\frac{1}{4}(X_{12} \mp X_{21}) 
 +\frac{i}{2\sqrt{3}} (X_{11} \pm \frac{3}{4}X_{22})
\label{hpm}
\end{equation}
where
\begin{equation}
X_{i,j}=\sqrt{\beta}\pi z \{J_i (\beta z) N_j(z)-N_i(\beta z) J_j(z)\}\;\;.
\label{X}
\end{equation}
The matrix in eq. (\ref{transfer}) is called the transfer matrix of
 the junction, denoted by $T_s$, hereafter.
We assume here that evanescent waves can be neglected. (Assumption III)
It can be easily confirmed that $T_s$ 
 in  eq. (\ref{transfer}) satisfies the time reversal symmetry
 and unitarity ( See Appendix I).
  The parameter $z$ is related to the Fermi energy $E_F$ as follows.
When  $|k|$ is near zero, channel number is always two
 independent of the radius of the nanotubes.
But as $|k|$ increases, the channel number 
increases firstly in the thicker tube, when $|k|$ exceeds $k_c=2\pi/R_5$ .
Owing to the linear dispersion relation eq. (\ref{dispersion}), 
$z=2 \pi k/k_c= 2 \pi E_F/E_c$ holds where
$E_c$ is the threshold Fermi energy corresponding to $k_c$.
The transmission rates are calculated from the transfer matrix,
and the conductance $\sigma$ is obtained by Landauer's formula
 as 
\begin{equation}
\sigma=2/|h_+|^2=\frac{24}{
\{\sum_{i=1}^2\sum_{j=1}^2 
(3/4)^{i+j-2}X_{i,j}^2 
 \} + 6 }\;\;.
\label{sigma}
\end{equation}
The obtained conductance has a remarkable feature that it does not
 depend on the angle $\phi$.
It is consistent with the scaling law with the
 two parameters $E_F/E_c$ and $\beta=R_7/R_5$ in Ref.\cite{tamurajunction2}.
Fig. \ref{efconductance}  shows the conductances calculated by the tight binding model,
 and those by eq. (\ref{sigma}).
Agreement between the two methods is fairly good.
When $\beta \sim 1$, $\sigma$ is almost constant 
 with a value  near  2 in units of $2e^2/h$.
As $\beta$ decreases, peak structures appear at $E$ slightly below $E_c$.
The former case, $\beta \sim 1$,  is treated mainly in the following sections.
In this case, $\sigma$ is almost the same as that of $E_F=0$, 
 which is 
\begin{equation}
\sigma=8/(\beta^3 + \beta^{-3}+2)\;\;.
\label{zerosigma}
\end{equation}
Eq. (\ref{zerosigma}) reproduces well  the numerical results
 in Ref.\cite{tamurajunction1}.

\section{Band structures of the periodic nanotube junctions}
From now on, $a$, which is the length of the translation vectors 
$\vec{e}_1,\vec{e}_2$, and $\frac{\sqrt{3}}{2}\gamma$, which is the hopping integral between the nearest  neighbors multiplied by $\frac{\sqrt{3}}{2}$, are taken to be units of the length and the energy, respectively.
So $a$ and $\frac{\sqrt{3}}{2}\gamma$ are omitted in the following expressions.
Fig. \ref{perioddevelop} shows the unit cell of the periodic junctions, which
  is composed by the two equivalent nanotube junctions, where 
  one of the junctions is rotated by $\pi$ with respect to
  the other.
The unit cell in the development map is determined by four vectors:
the circumference of the thicker tube $\vec{R_5}=(m_5,n_5)$, that of
 the thinner tube $\vec{R_7}=(m_7,n_7)$ , the vector connecting 
the two pentagons in the thicker tube part, 
 $\vec{L^{(5)}}=(L_1^{(5)},L_2^{(5)})$ and 
 that connecting the two heptagons in the thinner tube part 
 $\vec{L^{(7)}}=(L_1^{(7)},L_2^{(7)})$.
The components of these vectors are referred
  to $(\vec{e}_1,\vec{e}_2)$,
 so that all of them are integers.
In our discussion they can be taken arbitrary integers, so far as
 both of $m_7-n_7$ and $m_5-n_5$ are multiples of three.
The transmission matrix $T_s$ is determined by eq. (\ref{transfer})$\sim$ eq. (\ref{X}), using $R_j=|\vec{R_j}|=\sqrt{m_j^2+n_j^2+m_jn_j}\;\;(j=5,7)$, and $\cos \phi =(m_5m_7+n_5n_7+ \frac{1}{2}(m_5n_7+m_7n_5))/(R_5R_7)$.
The transfer matrix of the unit cell $T_{p}$ is obtained by combining the two identical transfer matrices of the junction, $T_s$.
Fig. \ref{combination}  is a schematic view of this combination.
The scattering occurs as $\vec{\alpha}^L_7=T_s\vec{\alpha}^L_5$ in the left
 junction and $\vec{\alpha}^R_7=T_s\vec{\alpha}^R_5$ in the right junction.
Here superscripts $L$ and $R$ mean the  the coordinate
 systems, origins of which are the left heptagon and the right heptagon
, respectively. 
The coordinate system $L$ is transformed to $R$ with the rotation by 
 $\pi$ and the translation $\vec{L^{(7)}}$.
Therefore the coordinates $\vec{q^L}$ and the sublattices $(A,B)$ of $L$ are transformed to those of $R$ as
\begin{eqnarray}
\vec{q^R} &= &\vec{L^{(7)}}-\vec{q^L} \nonumber \\
(\psi^R_{A}(\vec{q^R}),\psi^R_{B}(\vec{q^R}))
 & = &(\psi^L_{B}(\vec{q^L}),\psi_{A}^L(\vec{q^L})) 
\label{transform}
\end{eqnarray}
As is known from eq. (\ref{propK}) and eq. (\ref{propK'}),
 the Bloch state wave function corresponding to $\alpha_{7+}^{LK}$ and 
$\alpha_{7+}^{LK'}$ is
\begin{equation}
(\psi^L_{A}(\vec{q^L}),\psi_{B}^L(\vec{q^L}))=
(e^{\mp i \eta/2},\pm e^{\pm i\eta/2})
e^{i\vec{k'_7} \cdot \vec{q^L}}w ^{\pm(q^L_1-q^L_2)}\alpha_{7+}^{L}\;\;,
\label{a+}
\end{equation}
and that corresponding to $\alpha_{7-}^{RK}$ and $\alpha_{7-}^{RK'}$ is
\begin{equation}
(\psi^R_{A}(\vec{q^R}),\psi^R_{B}(\vec{q^R}))=
(e^{\mp i\eta/2},\mp e^{\pm i\eta/2})
e^{-i\vec{k'_7} \cdot \vec{q^R}}w ^{\pm(q^R_1-q^R_2)} \alpha_{7-}^{R}
\;\;,
\label{a-}
\end{equation}
where the upper signs and the lower signs correspond to $K$ and $K'$, respectively.
From eq. (\ref{transform}),eq. (\ref{a+}) and eq. (\ref{a-}),
  the relation between $\vec{\alpha^L}_{7+}$
 and $\vec{\alpha^R}_{7-}$ is obtained as
\begin{equation}
\left(\begin{array}{c}
  \alpha^{RK'}_{7-} \\
  \alpha^{RK}_{7-} \\
\end{array}
\right)
 = \exp(i \vec{k'_7} \cdot \vec{L^{(7)}})
\left( \begin{array}{cc}
w^{l_7},& 0\\
0 ,& -w^{-l_7} \\
\end{array} \right)
\left(\begin{array}{c}
  \alpha^{LK}_{7+} \\
  \alpha^{LK'}_{7+} \\
\end{array}
\right)
\label{lamda7}
\end{equation}
where $l_7\equiv L_1^{(7)}-L_2^{(7)}$.
In the same way,
\begin{equation}
\left(\begin{array}{c}
  \alpha^{LK}_{5+} \\
  \alpha^{LK'}_{5+} \\
\end{array}
\right)
 = \exp(i \vec{k'_5} \cdot \vec{L^{(5)}})
\left( \begin{array}{cc}
w^{l_5},& 0\\
0 ,& -w^{-l_5} \\
\end{array} \right)
\left(\begin{array}{c}
  \alpha^{RK'}_{5-} \\
  \alpha^{RK}_{5-} \\
\end{array}
\right)
\label{lamda5}
\end{equation}
where $l_5 \equiv L_1^{(5)}-L_2^{(5)}$.
 Since $w^3=1$, $l_i$ can take either values of  0, or $\pm1 \pmod{3}$.
In order to visualize the integer $l_i$,
 the Kekule pattern  with thick bonds and thin bonds
 are drawn in the development map 
so that the pentagons and the heptagons have only thin bonds
 as shown in Fig. \ref{perioddevelop}. Domain boundary caused
 by mismatch of the pattern, 
 which is called 'the phason line'
 according to Ref.\cite{akagi}, 
 occurs in the thicker tube (thinner tube) only when
 $l_5=\pm1$ ($l_7=\pm1$).\cite{note}
 The reason for this correspondence is 
 that the Bloch states at the $K$ and $K'$ corner points have
 the same periodicity as that of the Kekule pattern.
 Note that we do not intend here 
 that the Kekule pattern represents the bond alternation.
 In this paper, the Kekule pattern is used 
 only to show the periodicity of the Bloch states
 at the $K$ and $K'$ corner points.

The $2 \times 2$ diagonal matrices in equations
(\ref{lamda7}) and (\ref{lamda5}) are denoted by $\Lambda_5$
 and $\Lambda_7$, respectively.
Then the transfer matrix of the unit cell, $T_{p}$, is 
obtained as
\begin{equation}
T_{p} =
\left( \begin{array}{cc}
 T_1 ,& T_2^*\\
 T_2 ,& T_1^*\\
\end{array} \right)
\label{periodictransfer}
\end{equation}
where
\begin{eqnarray}
T_1=\Lambda_5^{1/2}
(^tt_1\Lambda_7t_1- ^tt_2\Lambda_7^{-1}t_2)\Lambda_5^{1/2}\;\;,\nonumber \\
T_2=\Lambda_5^{-1/2}(t_1^{\dag}\Lambda_7^{-1}t_2-t_2^{\dag}\Lambda_7t_1)
\Lambda_5^{1/2} \;\;.
\label{T12}
\end{eqnarray}
Above equations (\ref{T12}) show that $^tT_1=T_1$, and $^tT_2=-T_2^*$.
It means that
\begin{equation}
T_{p}^{-1}=T_{p}^*
\label{T-1}
\end{equation}
 where eq. (\ref{t-1}) in  Appendix is used.
If $\vec{x}$ is an eigen vector of $T_{p}$ with an eigen value $\lambda$,
 i.e., $T_{p}\vec{x}=\lambda \vec{x}$, 
 $\vec{x}$ is also the eigen vector of  
$(T_{p}+T_{p}^{-1})/2= {\rm Re}(T_{p})$ with the eigen value 
 $\frac{1}{2}(\lambda+1/\lambda)$.
 When $\lambda=\exp (i k^{(p)})$ with a real value of $k^{(p)}$,
 $k^{(p)}$ is the Bloch wavenumber of the periodic junctions.
 The Bloch wavenumber $k^{(p)}$ is obtained from the eigen value
 of the real matrix ${\rm Re}(T_{p})$ as $\cos( k^{(p)}) = 
\frac{1}{2}(\lambda+1/\lambda)$.
Furthermore ${\rm Re}(T_{p})$ can be
block diagonalized as
\begin{eqnarray}
\frac{1}{2}
\left( \begin{array}{cc}
 1 ,& 1\\
 1 ,& -1\\
\end{array} \right)
\left( \begin{array}{cc}
 {\rm Re}(T_1) ,& {\rm Re}(T_2)\\
 {\rm Re}(T_2) ,& {\rm Re}(T_1)\\
\end{array} \right)
\left( \begin{array}{cc}
 1 ,& 1\\
 1 ,& -1\\
\end{array} \right)
= \nonumber \\
\left( \begin{array}{cc}
 {\rm Re}(T_1+T_2) ,& 0\\
  0 ,&  {\rm Re}(T_1-T_2)\\
\end{array} \right)
\label{ReT}
\end{eqnarray}
The eigen values of ${\rm Re}(T_{p})$ can be represented
 by those of ${\rm Re}(T_1+T_2)$ or those of ${\rm Re}(T_1-T_2)$.
After all, the dimension of the matrix which has to be diagonalized
 can be made half.
Since in this paper the Fermi energy region is considered 
 where the channel number is kept to two,
  the dimension of ${\rm Re}( T_1+ T_2)$ is two.
The two energy bands  of the periodic junctions, $k^{(p)}_{+}$ and  $k^{(p)}_{-}$,  are obtained as
\begin{equation}
\cos (k^{(p)}_{\pm})= 
\frac{1}{2}\left( x_{11}+x_{22} \pm\sqrt{(x_{11}-x_{22})^2
+4x_{21}^2-4y_{21}^2} \right)
\label{periodicdispersion}
\end{equation}
 where $x_{ij}={\rm Re}(T_1)_{ij}$ and
$y_{ij}={\rm Re}(T_2)_{ij}$. 

We have to discuss $\Lambda$ and $T_s$ further
 to obtain the dispersion relation.
As for the phases of $\Lambda$,
 $\vec{k'_i} \cdot  \vec{L^{(i)}}= E\tilde{L_i}$,
  where $\tilde{L_i}$ is  the length of $\vec{L^{(i)}}$ measured
 along each tube axis direction,  i.e. , $\tilde{L_i}\equiv 
|\vec{L^{(i)}}\times\vec{R_i}|/|\vec{R_i}|\;\; (i=5,7)$.
It is because  
 $\vec{k'_5} \cdot \vec{R_5}=0$ and $\vec{k'_7} \cdot \vec{R_7}=0$,
 that is to say, $\vec{k'_5}$ and $\vec{k'_7}$ are parallel to
 each tube axis for the metallic nanotubes.
 For the discussion of $T_s$,
 the phases $p_{\pm}$ of $h_{\pm}$ in eq. (\ref{hpm}) 
 are defined by
 $h_+=\sqrt{1/T}\exp (ip_+)$
and  $h_-=-\sqrt{R/T}\exp (ip_-)$.
As seen from eq. (\ref{sigma}) and unitarity, 
 $T$ and $R$ are the transmission rate
 and the reflection rate per channel, respectively.
$T$ determines the Landauer's formula conductance $\sigma$, as
 $\sigma=2T$ in units of $2e^2/h$.
When the tubes have almost the same radius, $0.7R_5 < R_7  < R_5$, 
 expressions about $T$ and $p_{\pm}$ become simple, so
 the discussions  in the followings
  are concentrated on
 this case.
 In this case, the transmission rate $T$ is almost
 constant value irrespective of $z=2\pi E/E_c$.
  Therefore  eq. (\ref{zerosigma})
 is used instead of eq. (\ref{sigma}) in the following discussions.
The transmission coefficient $T$ is almost constant near unity,  i.e. , 
 $T \sim 1$ and $R \ll 1$.
When $1> R_7/R_5 >0.7$,  the range of $T$ is confined in $1.0 > T >0.76$.
 Fig. \ref{ppm} shows the quantities that $\{p_{\pm}(z_{j+1})-p_{\pm}(z_j)\}/(\Delta z)$ vs. $R_7/R_5$, where  $ \Delta z =\pi/5,\; z_j=j\Delta z, \;j=0,1,\cdots,9$.
 They represent $\frac{\partial p_{\pm}}{\partial z}$,
 for $ 0 < z < 2\pi$, i.e., $ |E| < E_c \;\; ( z= 2\pi E/E_c=R_5 E)$.
When $R_7/R_5 \ge 0.7$, the ten plots  of each $R_7/R_5$
  show almost the same value.
It means that the phases $p_{\pm}$ are almost proportional to $E$.
Fig. \ref{ppm} shows that the gradients of $p_+$  are very close to
 $1 - R_7/R_5$, while those of $p_-$, denoted by $g$,  
 are almost constant between $-0.1$ and $-0.05$.
 Therefore  when $R_7/R_5 > 0.7$ and $ |E|< E_c$, 
 approximate expressions can be obtained as
\begin{eqnarray}
p_1 &\sim &(R_5-R_7)E \;\;,  \nonumber \\
p_2 &\sim & gR_5 E \;\; ( g= -0.1 \sim -0.05).
\label{p1p2}
\end{eqnarray}

The following two combinations of the gradients of phases 
 determine the band structures.
One is $\Omega_+ \equiv \tilde{L_5}+\tilde{L_7}+2R_5-2R_7$
 and the other is $\Omega_- \equiv \tilde{L_5}-\tilde{L_7}+2gR_5$.
The former is the length of the unit cell, 
which is defined as the sum of the four 
 lengths, two of which are those of the tubes measured
 along each tube axis and the other two are those of the junctions
 measured  along the 'radial' direction.

 When there is no phason line in the thinner tube, $l_7=0$, 
  we obtain
\begin{equation}
\cos (k^{(p)}_{\pm})=\frac{1}{T}\cos(\Omega_+ E\pm\frac{2}{3}\pi l_5)
 -\frac{R}{T}\cos (\Omega_- E\pm\frac{2}{3}\pi l_5)\;\;.
\label{l7}
\end{equation}
On the other hand, when there is no phason line in the thicker tube, $l_5=0$,
 the following is obtained,
\begin{equation}
\cos (k^{(p)}_{\pm})=\frac{1}{T}\cos(\Omega_+ E\pm\frac{2}{3}\pi l_7)
 -\frac{R}{T}\cos (\Omega_- E \mp\frac{2}{3}\pi l_7)\;\;.
\label{l5}
\end{equation}
Fig.  \ref{nophason}  shows the comparison of the band structures calculated by equations (\ref{l7}), (\ref{l5}) and the tight binding model for  case (a) 
 where there is no phason line, i.e., $l_5=l_7=0$, and for  case
 (b) where there are the phason lines in only one side of the tubes, i.e.,
  $l_5=\pm1, l_7=0$ or  $l_5=0, l_7=\pm1$. 
 The Fermi level comes at the highest energy of the negative energy bands.
 \cite{fermi}
The results by the two different methods agree fairly  well with each other,
 especially for the two bands nearest to the Fermi level.
The two bands are degenerate in  case (a),
 while they cross near  $k^{(p)} = 2\pi/3$ in  case (b).
The latter shows that the system becomes metallic in case (b).
But there is a significant difference between the results by the two methods
 in case (a).
 The gap appears at $E=0$ in the tight binding model, while
it does not appear in eq.  (\ref{l7}) and eq.  (\ref{l5}).
 Because existence of this gap determines whether
 the system is metallic or semiconducting,
 this difference  can not be neglected.

To investigate the origin of this difference, 
we have to compare the transfer matrix obtained by the tight binding model
 with the one obtained by eq. (\ref{transfer})$\sim$ eq. (\ref{X}).
For the transmission rate $T$, agreement between the two methods is good,
 as is already seen in the preceding section.
The ratios between the matrix elements of $t_1$ and $t_2$ is almost
 the same in the two method, as will be discussed in the next section.
Therefore the cause of this difference must exist in the phase factors,
$p_{\pm}$.
We do not get complete explanations about this discrepancy  yet. 
 But some features  can be found as below.
The phase factors obtained by the tight binding model are written
as $p_+=\chi_+(R_5-R_7)E + \epsilon_+$ and 
 $p_-=\chi_- g R_5E + \epsilon_-$, where $\chi_{\pm}$ and $\epsilon_{\pm}$
 represent the difference from eq. (\ref{p1p2}).
Fig. \ref{grad} shows $\chi_+$ as a function of the absolute value of
 the angle, $|\phi|$.
The number of calculated junctions is 274 as explained
in the figure caption. 
The values of $\chi_+$ is almost unity in most of the cases, but
 it increases when $R_7/R_5$ approaches unity and $|\phi|$
 is near $\pi/6$.
Though it causes a change of the gradient of the dispersion,
 it does not affect existence of the gap.
The range of the ratio $\chi_-$ is about $ 0 < \alpha_- < 1$,
 and it does not affect the existence of the gap, either.
Fig. \ref{intercept} shows intercepts $\epsilon_{\pm}$ as a function of $R_5-R_7$.
The intercept of $p_+$ is almost constant with the values near 0.03$\pi$, 
 but that of $p_-$, denoted by $\epsilon_-$,
 approaches zero as $R_5-R_7$ increases.
Thus we speculate that the nonzero intercept $\epsilon_-$ comes from the
 effects of discreteness of the lattice and it
 causes the gap as is shown bellow.

The second term in eq. (\ref{l7}) and eq. (\ref{l5}) 
 can be considered almost constant  with respect to $E$
 compared to the first term,
 when $\Omega_+ \gg \Omega_-$, which is valid for the periodic junctions 
 with $ \tilde{L_7},\tilde{L_5} \gg |g|R_5$.
 Therefore  the argument in the second cosine term can be substituted
 with $2\epsilon_-$.
Then one can see that the r. h. s. of eq. (\ref{l7}) and eq. (\ref{l5}) 
 become larger than unity near $E=0$, and the gap opens at $k^{(p)}=0$.
The width of the gap $W_g$ is estimated as
\begin{equation}
W_g \sim 2 \arccos (T + R \cos( 2\epsilon_-))/\Omega_+
\label{gap0}
\end{equation}
If $\epsilon_-$ is common, $W_g$ increases as $T$ decreases,
and is in inverse proportion to the length of the unit cell, $\Omega_+$.

Next, the periodic junctions in which the both kinds of tubes have the phason lines is discussed.
The discussions only for $l_5l_7=1$ are necessary,
 since those for $l_5l_7=-1$ can be easily obtained
 from it  by substitution $\phi$ with $\pi/3-\phi$.
When $l_7l_5 =1$, the dispersion relation obtained by the effective mass theory
 is
\begin{eqnarray}
\cos (k^{(p)}_{\pm}) = \frac{1}{T}\cos(\Omega_+ E)\cos (\delta_+)
 -\frac{R}{T}\cos (\Omega_- E)\cos (\delta_-) \nonumber \\
 \pm\frac{1}{T}\sqrt{Y} 
\label{l5l7}
\end{eqnarray}
where
\begin{eqnarray}
 Y = \sin^2(\Omega_+ E)\sin^2(\delta_+)
 +R^2\sin^2(\Omega_- E)\sin^2(\delta_-) \nonumber \\
 -\frac{9}{8}R\sin^2(3\phi)(\cos\{(\Omega_+ - \Omega_-)E\}\cos\{(\Omega_+ 
+ \Omega_-)E\})+1) \;\;,
\label{Y}
\end{eqnarray}
\begin{equation}
\cos(\delta_{\pm})=\frac{1}{4} \mp \frac{3}{4}\cos(3\phi)\;\;.
\end{equation}

Fig. \ref{phason}  shows an example of the band structures calculated
 by the tight binding model and those calculated by eq. (\ref{l5l7}).
Agreement between the two methods is also satisfactory  in this case.
There are two types of the gap at $E=0$.
The gap opens  when $Y$ becomes negative and the r. h. s.
 in eq. (\ref{l5l7}) becomes a complex number.
It is denoted by type $I$.
The gap of the other type, denoted by $II$,
 occurs when the r. h. s. in eq. (\ref{l5l7})
 becomes larger than unity.
The gap of type $I$ appears at nonzero  $k^{(p)}$  while
 that of type $II$ appears at $\Gamma$ point, i.e., $k^{(p)}=0$,
 as is seen in Fig. \ref{phason} (a) and (b), respectively. 
 The band structures with the gap of type $II$
 are similar to those  for $l_5=l_7=0$ shown in Fig. \ref{nophason} (a).

By neglecting the $R^2$ term in $Y$ and expanding $Y$ with respect to $E$,
the width of the gap of type $I$, denoted by $W_g^I$, can be approximated as
\begin{equation}
W_g^I\simeq\frac{2\sin \left(\frac{3}{2}\phi \right)}{\Omega_+}
\sqrt{6 R (1+\cos^2(2\epsilon_-))/
(4-3\cos^2 \left(\frac{3}{2}\phi \right))}
\label{I}
\end{equation}
Due to the gap of type $I$,  the HOMO band and the LUMO band
 avoid each other when $k^{(p)} \neq 0, \pi$.
It is related to the symmetry as will be discussed in the next section.
When $\phi$ approaches zero, $W_g^I$ also approaches zero and
 the band structures become similar to those for $l_5=0,l_7=\pm1$ or
$l_7=0,l_5=\pm1$ which is shown in Fig.  \ref{nophason} (b).
As $\phi$ approaches $\pi/3$, the gap of type $II$ appears instead of 
 type $I$. Its width at $\phi=\pi/3$, denoted by $W_g^{II}$, is evaluated 
 for $-\pi < 2\epsilon_- < 0$ by
\begin{equation}
W_g^{II}\simeq 2 \arccos( T- R\cos(2\epsilon_-+\pi/3))/\Omega_+
\label{II}
\end{equation}
Here the correction caused by the tight binding model $\cos (2 \epsilon_-)$ 
 is included in eq. (\ref{I}) and eq. (\ref{II}).
 The gap width is again in inverse proportion to the length of
 the unit cell, $\Omega_+$.
By using Taylor expansion, $\arccos( 1 -x) \simeq \sqrt{2x}$ for $0< x \ll 1$,
 eq. (\ref{II}) can be approximated further as
\begin{equation}
W_g^{II} \simeq 2 \sqrt{R(2+2 \cos(2\epsilon_- + \pi/3))}/\Omega_+
\label{II-2}
\end{equation}
The common factor $\sqrt{R}$ in eq. (\ref{I}) and eq. (\ref{II-2})
 can be written by using eq. (\ref{zerosigma}) as 
$ \sqrt{R}=\sqrt{1-T} \simeq \frac{3}{2}(1-R_7/R_5)$.
 Thus relation between the gap width and the ratio of the circumferences
 $R_7/R_5$ is almost linear.
 When $\epsilon_-$ is near zero, eq. (\ref{I}) approaches eq. (\ref{II-2}) as
  $\phi$ approaches  $ \pi/3$. 
 It means that even when $\phi$ is close to $\pi/3$ so that type of the gap
 becomes $II$,
 eq. (\ref{I}) can be used approximately as the gap width.
In Fig. \ref{wg1}, the solid line  shows $W_g^{I}(\phi)/W_g^{I}(\pi/3)=2 
 \sin(3\phi/2)/\sqrt{4-3\cos^2(3\phi/2)} $ as a function of $\phi$.
It can be seen that the gap increases as $\phi$ approaches $\pi/3$.
The dotted line shows that for $l_5l_7=-1$, which is obtained from the
 solid line for $l_5l_7=1$ by reversing the horizontal axis.
 
\section{Discussions based on symmetries} \label{symmetry}

In this section, we discuss what can be said by considering only the
 symmetry of the junctions without solving the effective mass equations.
To discuss the symmetry, the scattering matrix $S$ is used.
Symmetric properties of $S$ and its relation to the transfer matrix 
 are summarized
 in the Appendix.
The scattering matrix $S$ determines the outgoing waves for the incoming waves
as
\begin{equation}
\left( \begin{array}{c} \vec{\alpha_{5-}} \\ \vec{\alpha_{7+}} \end{array}
\right)
= 
\left( \begin{array}{cc} r_5 ,&  t\\ ^tt ,& r_7 \end{array} \right),
\left( \begin{array}{c} \vec{\alpha_{5+}} \\ \vec{\alpha_{7-}}\end{array} \right)
\label{Smatrix}
\end{equation}
 where $\vec{\alpha_+}= ^t(\alpha_+^K,\alpha_+^{K'})$ and
$\vec{\alpha_-}= ^t(\alpha_+^{K'},\alpha_+^{K})$ , each component of which
 is defined in eq. (\ref{alpha}).
Note that the order of $K$ and $K'$ is reversed between $\vec{\alpha_+}$ and
$\vec{\alpha_-}$.
Consider the operation $Q_1$ defined as
 $Q_1(F_A^K,F_B^K,F_A^{K'},F_B^{K'})=(-F_B^{K'},F_A^{K'},F_B^K,-F_A^K)$.
 The amplitude $\vec{\alpha}$ is transformed by this operation $Q_1$ as
\begin{equation}
Q_1\vec{\alpha_{\pm}}=\pm\left(\begin{array} {cc} 0 ,&  1 \\ 1 ,& 0 \\
\end{array} \right)\vec{\alpha_{\pm}} 
\equiv \pm \sigma_1 \vec{\alpha_{\pm}} 
\label{q1}
\end{equation}
Since the effective mass theory equations (\ref{kpKB}),(\ref{kpKA}),
 (\ref{kpK'B}),(\ref{kpK'A}),
 and the boundary conditions
 (\ref{bound1}),(\ref{bound2}),(\ref{bound3}),(\ref{bound4})
 are invariant under the operation $Q_1$,
\begin{equation}
-\sigma_1 r_j \sigma_1 = r_j \;\;\;\;\; (j=5,7)
\label{sigma1r}
\end{equation}
\begin{equation}
\sigma_1 t \sigma_1 = t \;\;.
\label{sigma1t}
\end{equation}
Eq. (\ref{sigma1r}) means that $r_j$ is antisymmetric. On the other hand,
 $r_j$ is symmetric owing to time reversal symmetry.
So $r_j$ is diagonal.
From eq. (\ref{sigma1r}), eq. (\ref{sigma1t}) and unitarity of $S$,
one can write $r_j$ and $t$ as
\begin{equation}
r_j=\sqrt{R}e^{i\theta_j}\left(\begin{array}{cc} 1 ,& 0 \\0 ,& -1 \\
\end{array} \right)\;\;\;\;\;(j=5,7),
\label{q1r}
\end{equation}
and
\begin{equation}
t=\sqrt{T}e^{i(\theta_5+\theta_7)}\left(\begin{array}{cc} \cos(f) ,& i\sin(f)
 \\ i\sin(f) ,& \cos(f) \\
\end{array} \right)\;\;,
\label{q1t}
\end{equation}
where $R$ and $T=1-R$ are the reflection rate and the transmission
 rate, while $\theta_j$ and $f$ are certain real values.
The meaning of $f$ is discussed in the following.

Consider operation $Q_2$ shown in Fig. \ref{figq2} , where the thinner tube
 part is fixed, but the thicker tube part is rotated by $\pi/3$ in the
 development map.
 Under the operation $Q_2$, the upper development map
 is transformed into the lower development map, where
 the angle of the circumference of the 
 thicker tube increases by $\pi/3$ as $\eta'_5=\eta_5 +\pi/3$.
 Then the angle between the two tube axes defined by $\eta_7-\eta_5$
 decreases by $\pi/3$ as $\phi'=\phi-\pi/3$.
The two development maps in Fig. \ref{figq2} correspond to 
 an identical junction.\cite{note3}
The difference is only how to draw the cutting line on the honeycomb plane
 of the junction.
 Therefore the $S$ matrix of the upper development map becomes the same
 as that of the lower one  after an unitary transformation corresponding to
 the operation $Q_2$.
Following the same discussions as those for the boundary conditions (\ref{bound1}), 
(\ref{bound2}), (\ref{bound3})  and (\ref{bound4}), 
 $(F_A^K,F_B^K,F_A^{K'},F_B^{K'})|_{\theta+\pi/3}
=( \frac{1}{w} F_B^{K'},
w F_A^{K'}
,w F_B^K
,\frac{1}{w}F_A^K)|_{\theta}$.
By using it and $\eta'_5 = \eta_5+\pi/3$ in eq. (\ref{alpha}),
\begin{equation}
Q_2\vec{\alpha_{5\pm}}= \pm i \sigma_1 \vec{\alpha_{5\pm}} 
\label{q2}
\end{equation}
while $Q_2\vec{\alpha}_{7\pm}=\vec{\alpha}_{7\pm}$.
From this symmetry, one can derive
\begin{eqnarray}
r_5(\phi-\pi/3)&=& -\sigma_1 r_5(\phi) \sigma_1 \nonumber \\
t(\phi-\pi/3) &=& -i\sigma_1t(\phi) \;\;,
\label{rotsym}
\end{eqnarray}
which leads
\begin{equation}
f(\phi)=f(0)+\frac{3}{2}\phi \;\;.
\label{fphi}
\end{equation}
Lastly, we consider the coordinate transformation from the right-handed 
 coordinate to the left handed coordinate $Q_3$,
 $(x,y) \rightarrow (-x,y)$.
It causes 
$\eta_j \rightarrow -\eta_j$, $\phi \rightarrow -\phi$,
and exchange between the sublattices, $Q_3(F^K_A,F^K_B,F^{K'}_A,F^{K'}_B)=(F^K_B,F^K_A,F^{K'}_B,F^{K'}_A) $, so that
\begin{equation}
Q_3\vec{\alpha_{\pm}}=\left(\begin{array}{cc} 1 ,& 0 \\ 0,& -1 \end{array}
\right)\vec{\alpha_{\pm}} 
\equiv \sigma_2 \vec{\alpha_{\pm}} \;\;.
\label{q3}
\end{equation}
The result deduced from it is
\begin{eqnarray}
 r_j(-\phi) & = &\sigma_2 r_j(\phi) \sigma_2 \;\;\;\;\;(j=5,7)\nonumber \\
t(-\phi) &=& \sigma_2 t(\phi) \sigma_2 \;\;,
\label{q3rt}
\end{eqnarray}
which indicates that $f(0)=0$ in eq. (\ref{fphi}).
The transfer matrix obtained from the above discussions
 has the same form as eq. (\ref{t1}) and eq. (\ref{t2}).
Here it should be noted that these results can not be applied strictly
 to the scattering matrix when higher order terms of $k\cdot p$ are included in the effective mass equations, which make the equations to be variant under the operation $Q_1$.
Thus the invariance under the operation $Q_1$ holds within the linear
 approximation with $\vec{k'}$.
 Fig. \ref{t21}  show how the $Q_1$ invariance is accurate for
  the $S$ matrix calculated by the tight binding model.
 It shows values of ${\rm Im}(t_{21}/t_{11})$  
 as a functions of $\phi$. The range of the indices of junctions 
 $(m_5,n_5)$ - $(m_7,n_7)$ and that of the energy
 are the same as those of Fig. \ref{grad} and Fig. \ref{intercept}.
It can be seen that these plots coincide with $\tan (3\phi/2)$ quite well.
The other ratios between the matrix elements obtained  
 by the tight binding model also coincide with
 those of eq. (\ref{q1r}) and eq. (\ref{q1t}) fairly well:
$|\frac{t_{22}}{t_{11}}-1| < 0.1$, $|\frac{r_{22}}{r_{11}}+1| < 0.2$,
$|\frac{r_{21}}{r_{11}}| < 0.1$, and $|{\rm Re}(\frac{t_{21}}{t_{11}})|
|\frac{t_{11}}{t_{21}}| < 0.2$ for the 274 junctions in Fig. \ref{t21}.

The ratios between the matrix elements of the scattering matrix
 is determined in this way.
But as for the factors $h_{\pm}$, what can be known from the above discussions
is only that they are periodic even functions of $\phi$ with the
 period $\pi/3$.
One has to solve the effective mass equations to get more information
 on them.
But there is an important result obtained solely from the argument
 of the symmetries.
As is shown in the preceding sections, 
 the two bands $k_+^{(p)}$ and $k^{(p)}_-$
 are degenerate when neither of the two tubes has the phason line.
On the other hand, the two band avoid each other when
 $\phi \neq 0$ ( $\phi \neq \pi/3$ )
 and both of the two tubes have the phason lines $l_5l_7=1$ ($l_5l_7=-1$).
From eq. (\ref{periodicdispersion}), these can be explained  as below.
 The origin of the degeneracy in the former case is that
both of ${\rm Re}(T_1)$  and  ${\rm Re}(T_2)$ 
are diagonal and ${\rm Re}(T_1)_{11}={\rm Re}(T_1)_{22}$.
 The origin of the repulsion between the bands in the latter case is that
  ${\rm Re}(T_2)$ has nonzero off-diagonal elements. 
 These origins can be shown only  from the results
 in this section without solving
 the effective mass equations.

\section{Summary and conclusion}
The band structures of the periodic junctions composed by the two
 kinds of metallic  nanotubes are considered by combining the transfer
 matrix of the single junction, $T_s$.
 The transfer matrix $T_s$ is obtained in an analytical form
 by the effective mass equations with  assumptions I, II and III. The region of energy is $|E|<E_c$, where
 $E_c$ is the threshold energy above which more than two channels open.
 By combining the $T_s$'s,  the dispersion relation of the periodic junctions
 is also obtained analytically.
 Discussions are concentrated to 
 the case when the two tubes have almost the same radius, i.e.,
 $ 0.7 < R_7/R_5 < 1$.
 In this case, the transmission rate $T$ per channel 
 is near one and almost constant
 with respect to the energy.
 Agreement between the band structures by the effective mass theory
 and those by the tight binding model is satisfactory.
 Correspondence between the phason lines and the band structures near
 undoped Fermi level discussed in Ref.\cite{akagi}  appears naturally.
 When there is no phason line (case i), the degenerate bands appear. 
 When only one of the two kinds of tubes has the phason lines (case ii),
 the two bands cross with each other near $k=2\pi/3$.
 But in case (i), 
 there is a significant difference between the results of the two methods.
 The gap appears at $k=0$ in the tight binding model, while
 it does not appear in the effective mass equations. 
 The origin of this  difference is that the values of the phase factors $p_-$
 at $E=0$ are different  between the two methods.
 But explanation for it  is not enough yet.
 In these two cases (i) and (ii), the angle between the two tube axes, $\phi$,
  does not influence the band structures.
 But when both kinds of tubes have the phason lines, the band structures depend
 on $\phi$ as follows (case iii).
  When $l_5l_7=1$ ($l_5l_7=-1$),
  the gap width increases and the corresponding wave number
  changes from  $k^{(p)} \simeq 2\pi/3$ to $k^{(p)}=0$,
  as $\phi$ increases from $0$ to $\pi/3$ (decreases from $\pi/3$ to 0).
 The width of the bands and the gap near the undoped 
 Fermi level is in inverse proportion to the length of the unit cell 
  which is defined as the sum of the four 
 lengths, two of which are those of the tubes measured
 along each tube axis and the other two are those of the junctions
 measured  along the 'radial' direction, i.e., $R_5-R_7$.
 It is also found that the ratio between the matrix elements 
 of the transfer matrix $T_s$
 is determined only by the symmetries.
 By using only the symmetries and  without 
 solving the effective mass equations,  
 the degeneracy of the bands for case (i)
 and the gap at $k \neq 0$ for case (iii) can be derived. 
 
In this paper, it is found that the effective mass theory is very useful
 to analyze the electronic states near the undoped Fermi levels.
Though the discussions is limited to some type of
 the periodic nanotube junctions, similar phenomena have been found
 in other systems.
 One of the examples is another type of the periodic nanotubes composed by
 only one kind of the tube.\cite{akagi} Another example is
  pairs of disclinations in the monolayer
 graphite.\cite{discl}
  These  electronic states can be also classified according to
 the phason line pattern.
 We expect that similar discussions can be applied also to these cases.

\section*{Acknowledgment}
We wish to thank K. Akagi.
We would like to thank H. Matsumura and T. Ando for their useful
 suggestions.
This work has been supported by the Core Research for Evolutional 
 Science and Technology (CREST) of the Japan Science and Technology 
 Corporation(JST).

\section*{Appendix}
We consider the junction between  a left lead 1 and a right lead 2 
in the absence of magnetic field.
When there is no magnetic field, the Hamiltonian $H$ can 
be taken to be real $H^*=H$. 
In this case, if $\psi$ is a stationary state, $H\psi =E \psi$,
 $\psi^*$ is also a stationary state, $H\psi^* =E \psi^*$. 
The direction of the probability
 flow of $\psi^*$ is opposite to that of $\psi$.
In each lead, the propagating wave is represented
by $ \sum_{j=1}^{N_i} (x_{j+}^{(i)} \psi_j^{(i)} + x_{j-}^{(i)}\psi_j^{(i)*})$
 for the lead $i$ ($i=1,2$).
Here the $j'$th left going wave $\psi_j^{(i)*}$ 
is taken to be complex  conjugate of $j'$th right going  wave
 $\psi_j^{(i)}$.
They are normalized so that the probability flow 
is represented by $|\vec{x}^{(i)}_+|^2 - |\vec{x}^{(i)}_-|^2$ in the each lead.
The integer $N_i$ is the number of the extended state in each tube and
 called the channel number.
By solving the Schr$\ddot{\rm{o}}$dinger  equation, the outgoing wave 
$\vec{x}_{out}\equiv ^t(\vec{x}_{1-},\vec{x}_{2+})$
 is determined by the incoming wave $\vec{x}_{in}\equiv ^t(\vec{x}_{1+},\vec{x}_{2-})$
 by the scattering
 matrix $S$ as
\begin{equation}
\vec{x}_{out}= S \vec{x}_{in} \;\;,
\label{s1}
\end{equation} 
where
\begin{equation}
S=\left(\begin{array}{cc} r_1 ,& t \\
                    t' ,& r_2 \\
                 \end{array}
\right)
\label{blockS}
\end{equation}
In eq. (\ref{blockS}), $r_1$ and $r_2$ are an $N_1 \times N_1$ matrix
 and an $N_2 \times N_2$ matrix, respectively, and they show the reflection rate.
On the other hand, the block matrices $t$ and $t'$ show the transmission rate.
The conservation of the probability flow $|\vec{x}_{in}|^2=|\vec{x}_{out}|^2$
 guarantees that $S$ is unitary, that is to say,
\begin{equation}
S^{\dag}=S^{-1}\;\;.
\label{unitaryS}
\end{equation}
In the absence of the magnetic field, the complex conjugate
 of the wave function represented by eq. (\ref{s1}) is also a stationary state.
It is represented by $\vec{x}_{in}^*= S \vec{x}_{out}^*$, which means that
\begin{equation}
S^*=S^{-1}\;\;.
\label{timeS}
\end{equation}
From eq. ({\ref{unitaryS}) and eq. (\ref{timeS}), one can know
$S$ is a symmetry matrix $^tS=S$, which means that $r_1= ^t\!r_1$, $r_2= ^t\!r_2$ and
 $t'= ^t\!t$.
Here attention should paid when the base wave functions are unitary transformed,
 $\vec{x'}_{\pm}^{(i)}=U_{\pm}^{(i)} \vec{x}_{\pm}^{(i)}$.
For arbitrary unitary matrices $U_{\pm}^{(i)}$, eq. (\ref{unitaryS}) holds in the
 representation $\vec{x'}$.
In contrast to it, eq. (\ref{timeS}) holds only when $U_{-}^{(i)}=U_{+}^{(i)*}$.

When the two leads have the same channel number, i.e. , $N_1=N_2$, one can consider
the transfer matrix $T$ instead of the scattering matrix $S$.
By the transfer matrix, the propagating wave in the lead 2, 
  $\vec{x}_2\equiv  ^t(\vec{x}_{2+},\vec{x}_{2-})$,  is determined 
by that in the lead 1, $\vec{x}_1\equiv  ^t(\vec{x}_{1+},\vec{x}_{1-})$  as
\begin{equation}
\vec{x}_2= T \vec{x}_1\;\;,
\end{equation}
where
\begin{equation}
T=\left(\begin{array}{cc} t_{11} ,& t_{12} \\
                     t_{21},& t_{22}\\
                 \end{array}
\right)\;\;.
\label{blockT}
\end{equation}
From the time reversal symmetry, one can show
$t_{11}=t_{22}^* \equiv t_1$ and $t_{21}=t_{12}^* \equiv t_2$ by the
similar discussion to that about $S$.
Conservation of the probability flow is represented
by $t_{1}^{\dag}t_{1}-t_{2}^{\dag}t_{2}= 1$
and $^tt_{1}t_{2}-  ^tt_{2}t_{1}=0$.
From these, the inverse matrix of $T$ is
\begin{equation}
T^{-1}=\left(\begin{array}{cc} t_{1}^{\dag} ,& -t_{2}^{\dag} \\
                    - ^tt_{2},& ^tt_{1}\\
                 \end{array}
\right)\;\;.
\label{t-1}
\end{equation}
The relation between $S$ and $T$ is represented as
\begin{eqnarray}
t_{1}=(1/t)^* \nonumber \\
t_{2}=-(1/t)r_1
\label{st}
\end{eqnarray}

\begin{figure}
\caption{Development map of the nanotube.}
\label{tubetenkai}
\end{figure}

\begin{figure}
\caption{Development map of the nanotube junctions.
It is similar to that of Ref. \protect\cite{saitojunction}.
The lines '${\rm EP_7}$', '${\rm P_7P_5}$' ,'${\rm P_5G}$' 
 are connected and become identical with the lines 
'${\rm FQ_7}$','${\rm Q_7Q_5}$' and '${\rm Q_5H}$', respectively.
The rectangles '${\rm EP_7Q_7F}$' and '${\rm P_5GHQ_5}$' form the thinner tube and the thicker tube, respectively. 
'${\rm P_7P_5}$' is the rotated '${\rm Q_7Q_5}$' by angle of
  60 degrees and the quadrilateral '${\rm P_7P_5Q_5Q_7}$' forms 
 a junction part with a shape of a part of a cone. 
A heptagonal defect and a pentagonal defect are introduced
at ${\rm P_7(=Q_7)}$ and ${\rm P_5(=Q_5)}$, respectively.
Lines '${\rm EP_7}$' and '${\rm FQ_7}$' are parallel to the thinner tube
 axis and lines '${\rm P_5G}$' and '${\rm Q_5H}$' are  parallel to the thicker
 tube axis.
The direction of the circumferences of the tubes in the development map
 is represented by their angles $\eta_5$ and $\eta_7$ measured anti-clockwise
 with respect to the $x$ axis defined
 in Fig. \protect\ref{tubetenkai}.
 Thus the angle between the axes of the two tubes, $\phi$, is defined as
 $\phi=\eta_7-\eta_5$.}
\label{junctiontenkai}
\end{figure}

\begin{figure}
\caption{An example of  $\pi/3$ rotation. $B'_i$ and $A'_i$ indicates
 the $B$ lattice and $A$ lattice with their positions
$ 2i \vec{e_1} -(i+1) \vec{e_2}$ and $ (2i+1) \vec{e_1} -(i+1) \vec{e_2}$,
 respectively.
 They are aligned along the line $OQ$ which is parallel to the bonds.
 The angle between the line $OP$ and the line  $OQ$ is $\pi/3$.
The points $A_i$ and $B_i$ are aligned along the line $OP$ and their
 positions are $ i \vec{e_1}+i \vec{e_2}$.
By the rotation $\pi/3$ with respect to the point $O$, the points $B'_i$ and 
 $A'_i$ are transformed into the points $A_i$ and $B_i$, respectively.
}
\label{rotbound}
\end{figure}

\begin{figure}
\caption{Landauer's formula Conductances calculated by the tight binding model
 and those by the effective mass theory. 
 The former is shown by plots and the latter is shown by solid lines.
 It can be seen that agreement between them is good.
 The horizontal axis is 
the energy in units of the absolute value of
 the hopping integral, $|\gamma|=-\gamma$. 
The region of energy is
 $|E| < E_c \simeq 0.31|\gamma|$,  where the channel number is kept to two.
 The conductances of junctions connecting the $(2i+4,2i+1)$ tube
 and the (10,10) tube for $i=0,1,2,3$ are shown. 
 Values  of $R_7/R_5$ are attached to the corresponding plots.}
\label{efconductance}
\end{figure}

\begin{figure}
\caption{Development map of the periodic junction.
 The positions of the pentagonal defects and heptagonal defects
 are denoted by symbols 5 and 7, respectively.
 The circumferences of the thicker tube and the thinner tube 
 are represented by vectors $\vec{R_5}$ and $\vec{R_7}$.
 The upper bold line is connected with the lower bold lines
 so that the points connected by the circumference vectors
 become identical.
 Vectors connecting the two pentagonal defects and the two heptagonal defects
 in each tube part are denoted by $\vec{L^{(5)}}$ and $\vec{L^{(7)}}$, 
 respectively.
 The four vectors $\vec{R_5}$, $\vec{R_7}$, $\vec{L^{(5)}}$ and $\vec{L^{(7)}}$
 determine the bond network of the periodic junction treated in this paper
 uniquely. In this figure, $\vec{R_5}=(2,5)$, $\vec{R_7}=(1,4)$, 
 $\vec{L^{(5)}}=(4,-4)$ and  $\vec{L^{(7)}}=(3,-3)$. 
 In this figure,
 the thicker tube has the phason line represented by the dotted line
 where the Kekule pattern becomes incommensurate.
 An area between the two dotted lines 
 is the unit cell of this periodic junctions}
\label{perioddevelop}
\end{figure}

\begin{figure}
\caption{Schematic view of combination of two equivalent junctions
 to form the unit cell of the periodic junctions.}
\label{combination}
\end{figure}

\begin{figure}
\caption{ The gradients $\frac{\partial p_{\pm}}{\partial z}$,
 for $ 0 < z < 2\pi$, i.e., $ |E| < E_c \;\; ( z= 2\pi E/E_c=R_5 E)$.
 They are evaluated with $\{p_{\pm}(z_{j+1})-p_{\pm}(z_j)\}/(\Delta z)$, where  $ \Delta z =\pi/5, \;z_j=j\Delta z, \;j=0,1,\cdots,9$.
 The horizontal axis is the ratio of the circumferences, $R_7/R_5$.
 The gradients $\frac{\partial p_{+}}{\partial z}$
 are almost same as 
 $1 - R_7/R_5$ , while $\frac{\protect\partial p_{-}}{\partial z}$
 are almost constant between $-0.1$ and $-0.05$.
 As a reference, the solid line representing $1-R_7/R_5$ is shown.}
\label{ppm}
\end{figure}

\begin{figure}
\caption{The band structures of the periodic junctions for one of the
 two kinds of tubes has no phason line.
 The vertical axis is the energy in units of the absolute value of
 the hopping integral, $|\gamma|=-\gamma$. 
 (a)$\vec{R_7}=(1,10)$, $\vec{R_5}=(3,12)$, $\vec{L^{(7)}}=(4,-5)$
 and $\vec{L^{(5)}}=(6,-3)$. $R_7/R_5 \simeq 0.766$ and $\phi\simeq0.034\pi$.
 $l_5=l_7=0$, i.e., 
 neither of the two tubes has the phason line. 
 (b) $\vec{R_7}=(7,1)$, $\vec{R_5}=(6,3)$, $\vec{L^{(7)}}=(3,-2)$
 and $\vec{L^{(5)}}=(9,-3)$. $R_7/R_5 \simeq 0.951$ and $\phi\simeq0.070\pi$.
 $l_7=-1$ and $l_5=0$, i.e., 
 only the thinner tube has the phason line.}
\label{nophason}
\end{figure}

\begin{figure}
\caption{ Gradients $\protect\frac{\protect\partial p_+(z)}{\protect\partial z}$  at
 $z=0$ where $z= k R_5= 2E R_5/(\protect\sqrt{3} |\protect\gamma|)$ calculated by the tight binding model and  normalized by those  obtained by the effective mass theory, i.e., $(1-R_7/R_5)$.
They are denoted by $\chi_+$ and plotted for 274 junctions. 
 The horizontal axis is absolute values of the angle
 $|\phi|$ between the two tube axes in the development map in units of $\pi$.
 The range of the indices of the tubes,  $(m,n)$, 
 of the  calculated junctions
 is $ 1 \leq min(m,n) \leq 8$ and $|m-n|=3i,\;\;i=0 \sim 3$
   which satisfy $ 1 > R_7/R_5 > 0.7$.
The error bar indicates that maximum and minimum of the
 $\chi_+$ for each junction in the energy region $|E/\gamma|< 0.1$.
That is to say, it represents deviation from the linearity in this energy
 region.
To evaluate this error bar, 
the scattering matrices for 9 different energies equally spaced
 in this energy  region are calculated about each junction.
Open diamond plots, closed circle plots, and open square plots
 correspond to $1 > R_7/R_5 \geq 0.95$, $0.95> R_7/R_5 \geq 0.9$
 and $0.9 > R_7/R_5 \geq 0.7$, respectively.}
\label{grad}
\end{figure}

\begin{figure}
\caption{ Values of the phases
 $p_{\pm}$  at $E=0$ divided by $\pi$ 
 calculated by the tight binding model.
They correspond to the intercepts $\epsilon_{\pm}$ divided by $\pi$.
 In the effective mass theory, they are zero. The range of the calculated junctions is same as that of Fig. \protect\ref{grad}. 
The horizontal axis is
 difference between the circumference of thicker tube and that of the thinner tube, $R_5-R_7$, which corresponds to the length of the junction part.}
\label{intercept}
\end{figure}

\begin{figure}
\caption{The band structures of the periodic junctions for both of
 two kinds of tubes have the phason lines.
 The vertical axis is the energy in units of the absolute value of
 the hopping integral, $|\gamma|=-\gamma$. 
 (a)$\vec{R_7}=(10,1)$, $\vec{R_5}=(7,7)$,  $\vec{L^{(7)}}=(6,-5)$
 and $\vec{L^{(5)}}=(13,-6)$. $R_7/R_5 \simeq 0.869$ and $\phi\simeq0.14\pi$.
 $l_7=-1$ and $l_5=1$.
 (b) $\vec{R_7}=(6,6)$, $\vec{R_5}=(7,7)$,  $\vec{L^{(7)}}=(4,2)$
 and $\vec{L^{(5)}}=(10,-3)$. $R_7/R_5 \simeq 0.857$ and $\phi=0$
 $l_7=-1$ and $l_5=1$.}
\label{phason}
\end{figure}

\begin{figure}
\caption{The dependence of the gap width on the angle $\phi$
 scaled by its maximum value when both of the two
 kinds of tubes have the phason lines. 
 It is approximately represented
 by $2 \sin(3\phi/2)/\protect\sqrt{4-3\cos^2(3\phi/2)}$ as discussed in the text.
 The solid line and the dotted line correspond to the
 cases $l_5l_7=1$ and $l_5l_7=-1$, respectively.}
\label{wg1}
\end{figure}

\begin{figure}
\caption{The operation $Q_2$ which fixes the thinner tube part but
 rotates the thicker tube part by $\pi/3$ in the development map.
 The upper development map is transformed into the lower one
 under the operation $Q_2$.
 These two development map correspond to the identical junction, denoted
 by the (2,2) - (2,5) junction.}
\label{figq2}
\end{figure}

\begin{figure}
\caption{Imaginary part of $t_{21}/t_{11}$ as a function of angle
 between the two tube axes $\phi$. The ranges of the junctions and energies
 are same as those of Fig. \protect\ref{grad}.
 They are fitted well by $\tan (\frac{3}{2}\phi)$ represented by the 
 solid line. }
\label{t21}
\end{figure}


\begin{references}
\bibitem{helixnote}Though the word 'helically coiled nanotubes'
is used in Ref. \protect\cite{akagi} , the periodic junctions have various
 shapes such as coiled forms, straight forms and zigzag forms. 
 Thus we use the word 'periodic nanotube junctions' in this
 paper.
\bibitem{note}
Since the junction between two metallic nanotubes is considered,
the phason line does not appear in the  junction part.
If one consider the  junction connecting a metallic nanotube and a semiconducting nanotube or that connecting two semiconducting nanotubes,
the phason line appears between the pentagon and the heptagon in the junction part.
\bibitem{fermi}
The position of the Fermi level can not be known
 by the effective mass theory because it 
 can not treat the bands far from $E=0$.
 It is the tight binding calculations that show that
 it comes at the highest 
 energy of the negative energy bands.
\bibitem{note3}
One to one correspondence between the junction and its development map
 holds when the range of $\phi$ is taken to be $\phi_0 \leq \phi < \phi_0+\pi/3$.
Otherwise there are multiple different development maps representing
a same junction, where $\phi$ is different by a multiple of $\pi/3$.
\end{references}
\end{document}